\documentclass[conference]{IEEEtran}
\usepackage{etex}
\usepackage{ifpdf}
\ifpdf
\usepackage[draft,pdfborder={0 0 0}]{hyperref}
\fi

\usepackage{cite}
\usepackage{graphicx}

\usepackage[cmex10]{amsmath}
\usepackage{nicefrac}
\usepackage{dsfont} 

\usepackage{bm}

\usepackage{array}
\usepackage[caption=false,font=footnotesize]{subfig}

\usepackage{tabularx}
\usepackage{multirow}
\usepackage{multicol}
\usepackage{dcolumn}
\usepackage{booktabs}

\usepackage{fixltx2e}
\usepackage{stfloats}

\usepackage{tikz,pgf}
\usepackage{pgfplots}
\usetikzlibrary{shapes,positioning,arrows,decorations.markings,fit,calc,patterns,spy}

\usepackage{glossaries}

\usepackage{flushend}

\newcommand{\infoSet}{\mathcal{A}}
\newcommand{\frozenSet}{\mathcal{A}^c}

\title{On the Tradeoff Between Accuracy and Complexity in Blind Detection of Polar Codes}

\author{\IEEEauthorblockN{
Pascal Giard, Alexios Balatsoukas-Stimming, and Andreas Burg}\vspace{2pt}
  \IEEEauthorblockA{Telecommunications Circuits Laboratory\\Ecole polytechnique f\'ed\'erale de Lausanne (EPFL), 1015 Lausanne VD, Switzerland\\Email: \{pascal.giard,alexios.balatsoukas,andreas.burg\}@epfl.ch}}

\begin{document}

\newacronym[plural=CRCs]{crc}{CRC}{cyclic redundancy check}
\newacronym{bler}{BLER}{block-error rate}
\newacronym{mdr}{MDR}{missed-detection rate}
\newacronym{sc}{SC}{successive-cancellation}
\newacronym{bp}{BP}{belief-propagation}
\newacronym[plural=LLRs,firstplural=log-likelihood ratios (LLRs)]{llr}{LLR}{log-likelihood-ratio}
\newacronym{fastssc}{fast-SSC}{fast simplified \gls{sc}}
\newacronym{scl}{SCL}{successive-cancellation list}
\newacronym{spc}{SPC}{single-parity-check}
\newacronym{ddg}{DDG}{data-dependency graph}
\newacronym{amc}{AMC}{adaptive modulation and coding}

\maketitle

\begin{abstract}
  Polar codes are a recent family of error-correcting codes with a number of desirable characteristics. Their disruptive nature is illustrated by their rapid adoption in the $5^{th}$-generation mobile-communication standard, where they are used to protect control messages. In this work, we describe a two-stage system tasked with identifying the location of control messages that consists of a detection and selection stage followed by a decoding one. The first stage spurs the need for polar-code detection algorithms with variable effort to balance complexity between the two stages. We illustrate this idea of variable effort for multiple detection algorithms aimed at the first stage. We propose three novel blind detection methods based on belief-propagation decoding inspired by early-stopping criteria. Then we show how their reliability improves with the number of decoding iterations to highlight the possible tradeoffs between accuracy and complexity. Additionally, we show similar tradeoffs for a detection method from previous work. In a setup where only one block encoded with the polar code of interest is present among many other blocks, our results notably show that, depending on the complexity budget, a variable number of undesirable blocks can be dismissed while achieving a \acrlong{mdr} in line with the block-error rate of a complex decoding algorithm.
  
\end{abstract}

\section{Introduction}
\label{sec:intro}
\Gls{amc} is an effective technique that is used by most modern communications systems in order to adapt the information-data rate of the system to the conditions of the wireless channel over which transmission takes place. The \gls{amc} information is transmitted over a dedicated control channel, which commonly uses a restricted number of modulation and coding combinations. However, in recent standards substantial amounts of control data have to be transmitted so that \gls{amc} techniques are also used for the control channel. To avoid using an additional control channel for the control channel, blind detection techniques are commonly employed to determine if control messages are present. The location of control messages along with their code parameters need to be blindly detected. The blind code detection task has been shown to be NP-hard in general~\cite{Valembois2001,Balatsoukas_ITW_2018}, so that heuristic algorithms are used in practice. For example, various methods have been proposed for the blind detection of Hamming and BCH codes~(e.g., \cite{Yardi2014,Chabot2007} and references therein), convolutional codes~(e.g., \cite{Cluzeau2009,Moosavi2011} and references therein), Turbo codes~(e.g., \cite{Debessu2012,Tillich2014} and references therein), and LDPC codes~(e.g., \cite{Xia2014,Yu2016} and references therein). However, so far, the blind detection of polar codes, which were recently included in the 5G standard, has received little attention~\cite{Condo2018, Condo2017, Giard_SIPS_2017}.

\begin{figure}
  \centering
  \includegraphics[width=0.9\columnwidth]{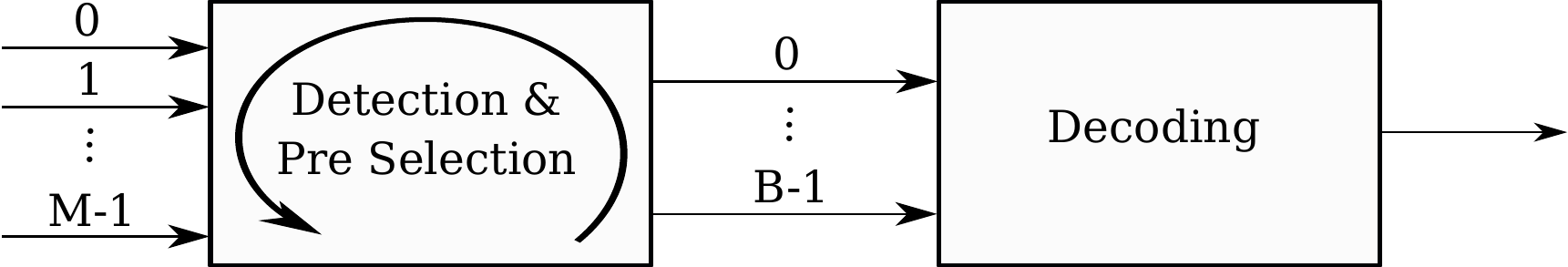}\vspace*{-0.5em}
  \caption{Overall system with detection stage and decoding stages.}
  \label{fig:detection-and-decoding}
\end{figure}

We are interested in the following blind detection scenario. We are given a set of $M$ received noisy blocks of length $N$, out of which exactly one originates from a codeword of a particular polar code $\mathcal{C}$, while the remaining blocks originate from random i.i.d. bit-sequences of length $N$ (that models data that is not of interest and encoded by some other code). The relevant polar codes used in 5G are concatenated with a \gls{crc}. Thus, we could attempt to decode all $M$ received blocks with a powerful decoding algorithm (e.g., \gls{scl} decoding) and to use the \gls{crc} to identify the block that was actually encoded using the polar code $\mathcal{C}$ with high probability. However, this approach has a high complexity and existing methods~\cite{Condo2018, Condo2017, Giard_SIPS_2017} consider a two-step approach, where a low-complexity algorithm is first used to discard some candidate blocks and only $B < M$ candidate blocks are passed on to the high-complexity decoding algorithm as illustrated in Fig.~\ref{fig:detection-and-decoding}.

\subsubsection*{Contributions}
In this paper, we present blind polar-code detection methods that enable the fine-grained adjustment of the tradeoff between the \gls{mdr} and complexity, where the complexity is determined by the effort spent on the detection-metric calculations and the number of retained candidates in the first and second stage, respectively. To this end, we present new polar-code detection algorithms based on \gls{bp} decoding of polar codes by adapting early-stopping criteria developed for \gls{bp} decoding for the purpose of detection. Using the newly proposed algorithms as well as a detection algorithm from \cite{Giard_SIPS_2017} (with a slight modification that enables complexity-performance tradeoffs), our simulation results show that good accuracy can be achieved with a significant reduction in complexity.

\subsubsection*{Outline}
The remainder of this paper starts with Section~\ref{sec:bg} that provides background on the construction, decoding, and detection of polar codes. In Section~\ref{sec:bpdet}, we describe three detection methods based on \gls{bp} decoding. Then, we explore the various tradeoffs between detection accuracy and complexity that can be achieved with the proposed algorithms in Section~\ref{sec:evaluation}. Finally, Section~\ref{sec:conclusion} concludes this paper.

\section{Background}
\label{sec:bg}
In this section, we provide some background on the construction of polar codes, and on  the \gls{sc}, \gls{fastssc}, and \gls{bp} decoding algorithms. We also briefly describe an existing \gls{fastssc}-based blind detection method for polar codes.

\subsection{Construction of Polar Codes}
Polar codes are linear block codes with a structured $N \times N$ generator matrix $\mathbf{G}$~\cite{Arikan2009}. Specifically, we have:
\begin{align}
	\mathbf{G}	& = \begin{bmatrix} 1 & 0 \\ 1 & 1 \end{bmatrix}^{\otimes n}, \label{eq:enc}
\end{align}
where $n = \log_2 N$, and $\otimes n$ denotes the $n$-fold Kronecker product. Let $\mathcal{A} \subseteq \{0,\hdots,N-1\}$ and let $\mathcal{A}^c$, referred to as the frozen-bit set, denote the complement of $\mathcal{A}$. Then, encoding of a $1 \times N$ polar codeword is performed as $\mathbf{x} = \mathbf{u}\mathbf{G}$, where $\mathbf{u}_{\mathcal{A}}$ contains information bits and $\mathbf{u}_{\mathcal{A}^c} = \mathbf{0}$. Due to the particular structure of the generator matrix defined in \eqref{eq:enc} and the associated decoding algorithm described in the following section, the different bit locations in $\mathbf{u}$ have different reliabilities. The set $\mathcal{A}$ for a polar code of rate $R = \frac{K}{N}$ is chosen to contain the $K$ most reliable bit positions~\cite{Arikan2009}.

\subsection{Decoding Algorithms}
\subsubsection{SC and Fast-SSC Decoding}
\Gls{sc} decoding traverses a \gls{ddg}~\cite{Arikan2009} and produces a vector of decision \glspl{llr} $\bm{\alpha}$ that is used to produce an estimate of $\mathbf{u}$, denoted by $\hat{\mathbf{u}}$, as follows:
\begin{align}
	\hat{u}_i = \delta_u(\alpha_i) = \left\{ 
		\begin{matrix}
			0, & \text{ if } i \in \frozenSet,\\
			0, & \text{ if } i \in \infoSet \text{ and } \alpha_i \geq 0,\\
			1, & \text{ if } i \in \infoSet \text{ and } \alpha_i < 0.
		\end{matrix}
		\right. \label{eq:deltau}
\end{align}
\Gls{fastssc} decoding reduces the decoding latency of \gls{sc} decoding by exploiting the structure of certain subgraphs in the \gls{ddg}. These subgraphs correspond to constituent sub-codes of the polar code that can be decoded efficiently (and often in a single time step) by specialized decoding algorithms~\cite{Sarkis_JSAC_2014}.

\subsubsection{BP Decoding}
BP decoding uses message passing on the factor graph defined by the generator matrix of a polar code. This factor graph has practically identical structure with the \gls{ddg} used for \gls{sc} decoding, with the main difference between the two algorithms being the scheduling of messages and the fact that in \gls{bp} decoding all messages are soft (i.e., \glspl{llr}) while a significant part of \gls{sc} decoding uses hard (i.e., $0$ or $1$) messages. \gls{bp} decoding at iteration $I$ consists of a right-to-left (or channel-to-information) pass which produces a vector of decision \glspl{llr} $\bm{\alpha}^I$ for the information vector $\mathbf{u}$, followed by a left-to-right (or information-to-channel) pass which produces a vector of decision \glspl{llr} $\bm{\beta}^I$ for the codeword vector $\mathbf{x}$. Similarly to \gls{sc} decoding, $\bm{\alpha}^I$ can be used to produce estimates for $\mathbf{u}$ using \eqref{eq:deltau}, while $\bm{\beta}^I$ can be used to produce estimates for $\mathbf{x}$ as follows:
\begin{align}
	\hat{x}^I_i = \delta_x(\beta^I_i) = \left\{ 
		\begin{matrix}
			0, & \text{ if } \beta^I_i \geq 0,\\
			1, & \text{ if } \beta^I_i < 0.
		\end{matrix}
		\right. \label{eq:deltax}
\end{align}
A detailed description of \gls{bp} decoding can be found in~\cite{Yuan2014}.

\subsection{Fast-SSC-Based Blind Detection of Polar Codes}
In this section, we briefly summarize an algorithm that we proposed in previous work \cite{Giard_SIPS_2017}. This method is based on \gls{fastssc} decoding \cite{Sarkis_JSAC_2014}, where a detection metric is calculated with update rules that exploit the inherent structure of the constituent codes that compose a polar code. In this \gls{fastssc}-based method, a detection metric $\mathcal{D}_t$ is initialized to $\mathcal{D}_0=0$ and is progressively updated as more and more leaf nodes in the decoder tree are visited. A threshold on the detection metric can then be set in order to decide whether a polar-encoded codeword is present or not. A straightforward way to adapt this method in order to obtain various tradeoffs between detection accuracy and calculation effort is to visit only a limited number of nodes in the decoder tree instead of traversing the entire tree, and this is the method we use to obtain the results in Section~\ref{sec:evaluation}.

In \cite{Giard_SIPS_2017}, detection-metric calculations are proposed for three constituent-code types: rate-0 codes, Repetition codes, and \gls{spc} codes. In this work, we initially used all three types but owing to its single parity bit, as opposed to the rate-0 and Repetition codes, \gls{spc} codes are the least reliable code of the three. As will be shown in Section~\ref{sec:evaluation}, it may be beneficial not to update the detection metric after visiting an \gls{spc} node. For a detailed description of the detection-metric calculations, we refer the reader to \cite{Giard_SIPS_2017}.

\section{BP-Based Blind Detection}
\label{sec:bpdet}
Existing blind detection methods for polar codes use variants of \gls{sc} decoding for detection~\cite{Condo2018, Condo2017, Giard_SIPS_2017}. Due to the stringent latency requirements for blind detection imposed by the 5G standard, blind detection methods based on \gls{bp} decoding are of interest for the first stage because \gls{bp} decoding is highly parallelizable and its complexity can be scaled easily by changing the number of performed iterations. Thus, in this section we propose three novel blind detection methods for polar codes based on \gls{bp} decoding that are inspired by early-stopping criteria found in the literature~\cite{Yuan2014,Ren2015,Abbas2015,Simsek2016,Zhang2017,Yu2018}. \Gls{bp} decoding refines the reliabilities of the \glspl{llr} at each iteration, a property that all proposed methods exploit.

\subsection{Method 1: Tracking of Decision-LLR Signs}
Under \gls{bp} decoding, decodable noisy polar-encoded blocks should converge to some codeword as the number of iterations increases. Thus, the signs of the decision \glspl{llr} are expected to gradually stabilize, meaning that an increasing number of decision \glspl{llr} should have the same signs between different iterations. This fact can be exploited in order to define the following metric for blind \gls{bp}-based detection of polar codes:
\begin{align}
  \mathcal{D}^{I}_{LS} & =  \left|\left\{i \in \{0,\hdots,N-1\}: \text{\small sign}\left(\alpha^{I}_i\right) = \text{\small sign}\left(\alpha^{I-1}_i\right) \right\}\right|,
  \label{eq:bp3}
\end{align}
where $\alpha^{I}_i$ is the decision \gls{llr} at bit location $i$ at iteration $I$, and $\alpha^{I-1}_i$ is the decision \gls{llr} for the same bit location but at iteration $(I-1)$. In words, $ \mathcal{D}^{I}_{LS}$ is the number of decision-\gls{llr} signs that remained unchanged in-between iteration $I$ and the one preceding it.

\subsection{Method 2: Frozen-Bit Set Inspection}
By construction of the polar code, we know that $\bm{u}_i=0$ for $i \in \frozenSet$. As such, even though the frozen channels are the least-reliable channels, one would expect that, as \gls{bp} decoding converges, an increasing number of decision \glspl{llr} corresponding to frozen-bit locations should become non-negative. Using this observation, we can define the following metric for blind \gls{bp}-based detection of polar codes:
\begin{align}
  \mathcal{D}^{I}_{FS} & = \left|\left\{i \in \frozenSet: \alpha^{I}_i \geq 0\right\}\right|,
  \label{eq:bp1}
\end{align}
where $\alpha^{I}_i$ is the decision \gls{llr} at bit location $i$ at iteration $I$. In words, $\mathcal{D}^{I}_{FS}$ is the number of non-negative frozen-bit decision-\gls{llr} signs at iteration $I$.

\subsection{Method 3: Decision-Vector Re-Encoding}
In \gls{bp} decoding, the right-to-left pass performs decoding, while the left-to-right pass is similar to soft encoding of the decision \glspl{llr} obtained with the right-to-left pass. For this reason, as \gls{bp} decoding converges, one would expect the re-encoded version of the decision vector $\hat{\bm{u}}^{I}$ to become equal to the decision vector $\hat{\bm{x}}^{I}$ obtained with \eqref{eq:deltax}, i.e., by taking hard decisions on the right-hand-side \glspl{llr} of the \gls{bp} graph. In other words, if decoding is successful, we should have $\hat{\mathbf{u}}^{I}\mathbf{G}=\hat{\mathbf{x}}^{I}$, and we can use this property in order to define the following metric for blind \gls{bp}-based detection of polar codes:
\begin{align}
  \mathcal{D}^{I}_{RE} & = \left|\left\{i \in \{0,\hdots,N-1\}: \tilde{x}^{I}_i = \hat{x}^{I}_i \right\}\right|,
\label{eq:bp2}
\end{align}
where $\tilde{\mathbf{x}}^{I}=\hat{\mathbf{u}}^{I}\mathbf{G}$. In words, $\mathcal{D}^{I}_{RE}$ is equal to the number of bit positions of $\tilde{x}^{I}_i$ and $\hat{x}^{I}_i$ that agree at iteration $I$.  

\section{Acurracy Vs Complexity}\label{sec:evaluation}

In this section, we quantify the accuracy of various detection methods by plotting the \glsfirst{mdr} as a function of the number of candidates $B$ passed to the second stage, i.e., to an \gls{scl} decoder. 

\subsection{System Setup}
Inspired by the LTE standard~\cite{3GPPLTER8} and by the discussions that took place at the 3GPP RAN1 meetings towards the creation of the next-generation mobile-communication standard (5G), we examine the performance for control messages that consist of a 16-bit identifier, an 8-bit payload, leading to a total of $K=24$ information bits, and a \gls{crc} of $C = 16$ bits. These control messages are encoded using a polar code of rate $R = \frac{K+C}{N}$, where $N = 256$. The actual information rate of this scheme is $R_{\text{inf}} = \frac{K}{N}$ and all $E_b/N_0$ values in the simulations are calculated using $R_{\text{inf}}$. 

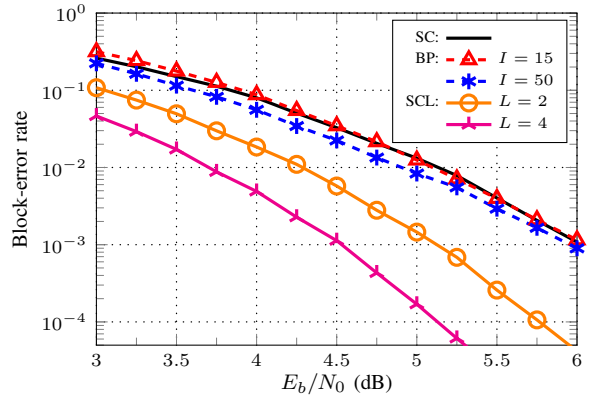
\begin{figure}[t]
  \centering
  \usetikzlibrary{plotmarks}

\definecolor{darkgreen}{RGB}{0, 128, 0}
\definecolor{mypurple}{RGB}{153, 71, 155}

\begin{tikzpicture}

  \pgfplotsset{
    grid style = {
      dash pattern = on 0.05mm off 1mm,
      line cap = round,
      black,
      line width = 0.5pt
    },
    title style = {font=\fontsize{7pt}{7.2}\selectfont},
    label style = {font=\fontsize{8pt}{7.2}\selectfont},
    tick label style = {font=\fontsize{7pt}{7.2}\selectfont}
  }

  \begin{semilogyaxis}[%
    xlabel=$E_b/N_0$ (dB),xtick={1,1.5,2,...,10},%
    xlabel style={yshift=0.8em},%
    minor x tick num={1},
    xmin=3.0,xmax=6.0,%
    ymin=5e-5,ymax=1e0,%
    ylabel=Block-error rate, ylabel style={yshift=-0.6em},%
    width=0.9\columnwidth, height=6.0cm, grid=major,%
    legend style={
      anchor={north east},
      cells={anchor=west},
      column sep=0mm,
      row sep=-0.5mm,
      font=\fontsize{6pt}{7.2}\selectfont,
      mark size=3.0pt,
      mark options=solid
    },
    legend columns=2,
    mark size=3.0pt,
    mark options=solid]
    
    \addlegendimage{empty legend}
    \addlegendentry[anchor=east]{SC:}
    \addplot[very thick,color=black] table[x=EbN0dB,y=FER] {data/256.24+16.SC.s0.95.csv};
    \addlegendentry{}

    \addlegendimage{empty legend}
    \addlegendentry[anchor=east]{BP:}
    \addplot[very thick,dashed,color=red,mark=triangle] table[x=EbN0dB,y=FER] {data/256.24+16.BP15.s0.95.csv};
    \addlegendentry{$I=15$}
    \addlegendimage{empty legend}
    \addlegendentry[anchor=east]{\phantom{BP:}}
    \addplot[very thick,dashed,color=blue,mark=asterisk] table[x=EbN0dB,y=FER] {data/256.24+16.BP50.s0.95.csv};
    \addlegendentry{$I=50$}

    \addlegendimage{empty legend}
    \addlegendentry[anchor=east]{SCL:}
    \addplot[very thick,color=orange, mark=o] table[x=EbN0dB,y=FER] {data/256.24+16.l2.s0.95.csv};
    \addlegendentry{$L=2$}
    \addlegendimage{empty legend}
    \addlegendentry[anchor=east]{\phantom{SCL:}}
    \addplot[very thick,color=magenta, mark=Mercedes star] table[x=EbN0dB,y=FER] {data/256.24+16.l4.s0.95.csv};
    \addlegendentry{$L=4$}

  \end{semilogyaxis}
\end{tikzpicture}\vspace*{-1.2em}
  \caption{Error-correction performance of a polar code with $N=256$, $K=24$, and $C=16$ under various decoding algorithms.}
  \label{fig:ec-perf}
\end{figure}

For reference, the \gls{bler} for a polar code with $N=256$, $K=24$, and $C=16$ decoded with the \gls{sc}, \gls{scl}, and \gls{bp} decoding algorithms is illustrated in Fig.~\ref{fig:ec-perf}. All simulation results are given for random codewords, BPSK modulation, and an AWGN channel. The 16-bit \gls{crc} used has the generator polynomial $z^{16}+z^{12}+z^5+1$. We use a scaled-min-sum version of the \gls{bp} decoding algorithm with a flooding schedule and a scaling factor of 0.9375. Results are for at least 50\,000 blocks and until a minimum of 500 blocks in error were found for each $E_b/N_0$ point. From Fig.~\ref{fig:ec-perf}, it can be seen that the \gls{bler} after 15 iterations of scaled-min-sum \gls{bp} decoding is nearly identical to that of \gls{sc} decoding. Increasing the number iterations from 15 to 50 provides a coding gain little under 0.25\,dB at a \gls{bler} of $10^{-2}$. At a \gls{bler} of $10^{-3}$, compared to the other simulated decoding algorithms, \gls{crc}-aided \gls{scl} decoding provides coding gains of approximately 0.8\,dB and 1.5\,dB, with list sizes $L\in\{2,4\}$, respectively.

The parameters mentioned above, used for the \gls{bler} simulations, also apply for the accuracy simulations, where the accuracy is expressed in terms of the \gls{mdr} as a function of the number of retained candidates. Furthermore, a set of $M=44$ blocks are generated per trial among which only one is encoded with the expected polar code while the remaining ones contain random i.i.d. bit-sequences, and simulations are for 100\,000 trials. The results are for an $E_b/N_0$ value that corresponds to a \gls{bler} of $10^{-2}$ (i.e., the beginning of the region of interest for wireless communications) under \gls{crc}-aided \gls{scl} decoding with a maximum list size $L=2$. We define a missed detection as the event where a block has been encoded with the expected polar code, its noisy realization is decodable, and the detection method failed to rank it among the $B$-best candidates (i.e., the candidates that would be passed to the decoder).

\begin{figure*}
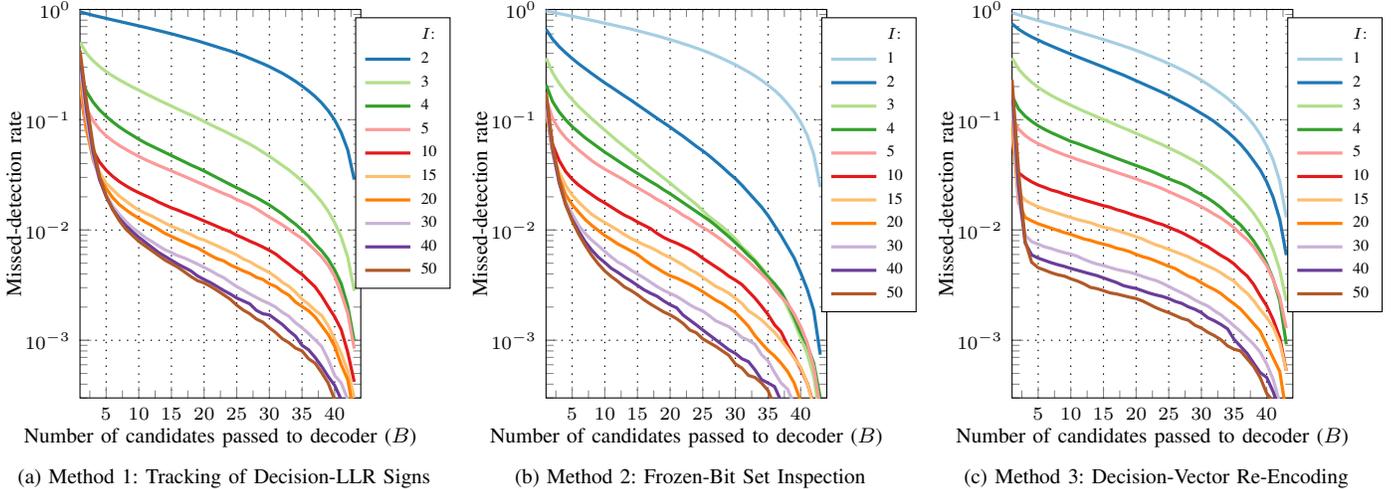

  \hspace*{-1.2em}
  \subfloat[Method 1: Tracking of Decision-LLR Signs]{
    \input{figures/mdr-256.24+16.EbN0.4.286.DetBP3.max50.100k}
    \label{fig:BP3:256.8+16+16}}\hspace*{-0.9em}%
  \subfloat[Method 2: Frozen-Bit Set Inspection]{
    \input{figures/mdr-256.24+16.EbN0.4.286.DetBP1.max50.100k}
    \label{fig:BP1:256.8+16+16}}\hspace*{-0.9em}%
  \subfloat[Method 3: Decision-Vector Re-Encoding]{
    \input{figures/mdr-256.24+16.EbN0.4.286.DetBP2.max50.100k}
    \label{fig:BP2:256.8+16+16}}
  \caption{BP-Based Detection: MDR as a function of the number of retained candidates $B$ for a polar code with $N=256$, $K=24$, and $C=16$.}\vspace*{-1.2em}
  \label{fig:BP:256.8+16+16}
\end{figure*}

\subsection{BP-Based Detection}
In Fig.~\ref{fig:BP:256.8+16+16}, we show the \gls{mdr} of the \gls{bp}-based methods, described Section~\ref{sec:bpdet}, as a function of the number of candidates passed to the decoding stage for various number of iterations. For these methods, the detection complexity of the first stage increases as a function of the number of decoding iterations $I$. It should be noted that method 1 is relying on the tracking of decision-\gls{llr} signs in-between two iterations, meaning that at least two decoding iterations are required. 

In Fig.~\ref{fig:BP:256.8+16+16}, we can observe that the number of performed \gls{bp} iterations greatly affects the number of candidates that need to be passed to the second stage in order to achieve a given \gls{mdr}. For example, at a fixed \gls{mdr} of $10^{-1}$, for method 1 the required number of retained candidates $B$ quickly drops from 40 to 20, from iterations $I=2$ and $I=3$, then to $M=6$ at iteration $I=4$. Furthermore, although the difference in terms of error-correction performance between 15 and 50 iterations of \gls{bp} decoding is small, as illustrated in Fig.~\ref{fig:ec-perf}, the effect on detection accuracy is significant. The most extreme case is with method 3, illustrated in Fig.~\ref{fig:BP2:256.8+16+16}, where it can be seen that for $B=4$ retained candidates, increasing $I$ from 15 to 50 reduces the \gls{mdr} from little under $2 \times 10^{-2}$ down to $4 \times 10^{-3}$.

To keep the complexity of the first stage low while achieving an \gls{mdr} of $10^{-2}$, the \gls{bp}-based methods were found to require $I=5$, $I=3$, and $I=7$ iterations for method 1, method 2, and method 3, respectively, to pass no more than $\nicefrac{3}{4}$ of the $M=44$ candidates (i.e., 33 candidates) to the high-complexity \gls{scl} decoder of the second stage. For the same \gls{mdr}, all three methods are shown to be capable of dismissing at least half of the candidates within $I=15$ iterations, where method 2 even manages to keep no more than $\nicefrac{1}{4}$ of the candidates. We also observe that, for all \gls{bp}-based methods, increasing the number of iterations leads to diminishing returns. Comparing the curves of methods 2 and 3, we note that their slopes greatly differ. This hints that these methods may have different use cases, e.g., for an \gls{mdr} of $10^{-2}$ method 2 is quickly capable of dismissing candidates whereas method 3 appears more suitable to iteratively reduce the \gls{mdr} at a fixed and small number of retained candidates.

\begin{figure*}
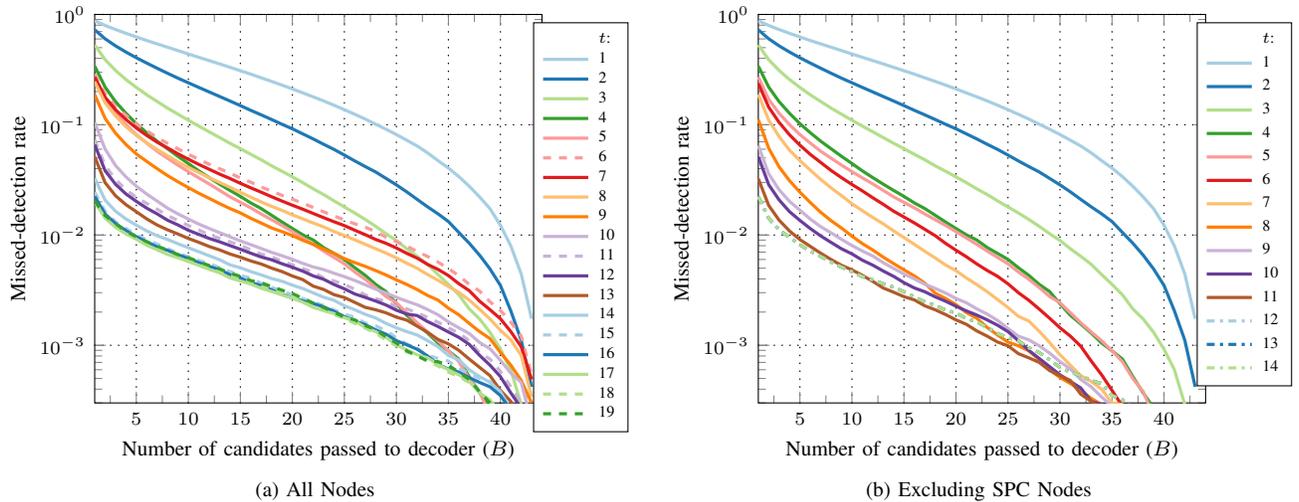

  \centering
  \subfloat[All Nodes]{
    \input{figures/mdr-256.24+16.EbN0.4.286.DetSiPS.100k}
    \label{fig:SiPSwSPC:256.8+16+16}}%
  \subfloat[Excluding SPC Nodes]{
    \input{figures/mdr-256.24+16.EbN0.4.286.DetSiPS.noSPC.100k}
    \label{fig:SiPSwoSPC:256.8+16+16}}%
  \caption{Fast-SSC-Based Detection: MDR as a function of the number of retained candidates $B$ for a polar code with $N=256$, $K=24$, and $C=16$.}\vspace*{-1.2em}
  \label{fig:SiPS:256.8+16+16}
\end{figure*}

\subsection{Fast-SSC-Based Detection}
In Fig.~\ref{fig:SiPS:256.8+16+16}, we show the \gls{mdr} of the \gls{fastssc}-based method, described in Section~\ref{sec:bg}, as a function of the number of candidates passed to the decoding stage for various number visited leaf nodes. For that method, the detection complexity of the first stage increases as a function of these visits. For readability, the number of visited leaf nodes $t$ in Fig.~\ref{fig:SiPS:256.8+16+16} is only increased when a leaf node used in the detection-metric calculations is visited, e.g., visiting a rate-1 node leaves the detection metric unchanged and is thus not accounted for in the value of $t$ shown in the plots.

Two plots for the \gls{fastssc}-based method are presented in Fig.~\ref{fig:SiPS:256.8+16+16}. The left-hand-side one, Fig.~\ref{fig:SiPSwSPC:256.8+16+16}, shows the accuracy of the detection metric when all update rules proposed in our previous work \cite{Giard_SIPS_2017} are used, where the dashed curves are the result of updates by a \gls{spc} node. The right-hand-side one, Fig.~\ref{fig:SiPSwoSPC:256.8+16+16}, shows the accuracy of the same method when \gls{spc} nodes are excluded, i.e., when \cite[Eq.\,(4)]{Giard_SIPS_2017} is not applied.

Looking at Fig.~\ref{fig:SiPSwSPC:256.8+16+16}, a certain number of crossovers between curves for different values of $t$ can be observed, hinting that at least one update rule may be detrimental to the accuracy. The first such crossover occurs when $t$ increases from 5 to 6. The update applied at $t=6$ corresponds to an \gls{spc} node. Fig.~\ref{fig:SiPSwoSPC:256.8+16+16} shows the accuracy of the same method where that update rule is excluded. From that figure it can be seen that all curves follow the same trajectory, there is no sudden change of slope anymore. Under \gls{sc}-based decoding, the \gls{bler} of that polar code at the $E_b/N_0$ of interest is approximately  $5 \times 10^{-2}$. Our hypothesis is that at such a high \gls{bler}, the \gls{spc} nodes are too unreliable to be beneficial for detection.

For a \gls{mdr} of $10^{-2}$, the \gls{fastssc}-based method is able to dismiss over a quarter of the $M=44$ candidates with a $t$ as small as 3. Excluding the \gls{spc} metric updates and visiting at least 12 of the 14 contributing nodes, 40 of the 44 candidates can be dismissed while achieving the same \gls{mdr}. Fig.~\ref{fig:SiPSwoSPC:256.8+16+16} also shows that a quarter of the candidates can be dismissed with a $10^{-3}$ \gls{mdr} by visiting at least 6 of the 14 the nodes that update the detection metric.

\subsection{Hardware Implementation Considerations}
The hardware-implementation complexity of all proposed methods is largely dominated by that of the underlying decoder. Comparing the complexity of \gls{bp}-based methods against that of the \gls{fastssc} one is challenging as the nature of their respective decoders can significantly vary. However, comparing the overhead of the \gls{bp}-based methods against each other, method 1 requires the storage of $N$ decision-\gls{llr} signs, and $N$ comparisons. Method 2, on the other hand, is the least complex, requiring only $(N-k)$ comparisons. Lastly, method 3 requires $N$ comparisons and additional circuitry to carry out the hard re-encoding step.

\section{Conclusion}\label{sec:conclusion}
In this paper, we described a two-stage system tasked with the location of control messages encoded with polar codes, where the first stage calculates a detection metric and selects a set of candidates to be passed to a second stage consisting of a high-complexity decoder.
The key to enable complexity-performance tradeoffs is to have a first stage which progressively computes and refines a metric which is then used for pre-selecting a set of candidates that should be passed on to the second stage. 
More steps in the first stage provide a better metric that allows for reliably excluding more candidates from the second stage, but they also add complexity to the first stage.
In this paper we presented multiple algorithms to implement the first stage, three of them are based on \glsfirst{bp} decoding, whereas the fourth is derived from \gls{fastssc} \cite{Giard_SIPS_2017}. In the first three methods, each step involves a refinement of the channel and codeword log-likelihood ratios using the \gls{bp} decoding algorithm followed by a simple metric calculation derived from early-termination criteria. For the fourth method, the metric itself is calculated and refined in steps based on intermediate results of the decoding algorithm presented in~\cite{Giard_SIPS_2017}.\vspace*{-0.6em}

\section*{Acknowledgement}
This work has been supported by the Swiss National Science Foundation under grant \href{http://p3.snf.ch/project-175813}{\#175813}.

\bibliographystyle{IEEEtran}
\bibliography{IEEEabrv,ConfAbrv,refs}

\end{document}